\documentclass[notitlepage,onecolumn,12pt,aps,nofootinbib,superscriptaddress,tightenlines]{revtex4-1}

\usepackage[colorlinks=true,citecolor=blue,urlcolor=blue,linkcolor=blue]{hyperref}
\usepackage{natbib}

\usepackage{amsmath,amsfonts,amssymb,amsthm,textcomp,mathrsfs,mathrsfs,bbm}
\usepackage{xfrac,braket}

\newtheorem{theorem}{Theorem}
\newtheorem{lemma}[theorem]{Lemma}

\newcommand{\beq}{\begin{equation}}
\newcommand{\enq}{\end{equation}}
\newcommand{\beqa}{\begin{eqnarray}}
\newcommand{\enqa}{\end{eqnarray}}
\newcommand{\beqan}{\begin{eqnarray*}}
\newcommand{\enqan}{\end{eqnarray*}}

\newcommand{\bel}{\begin{lemma}}
\newcommand{\enl}{\end{lemma}}
\newcommand{\bet}{\begin{theorem}}
\newcommand{\ent}{\end{theorem}}

\newcommand{\tr}{\mathrm{Tr}}

\begin {document}

\title{A gambling interpretation of some quantum information-theoretic quantities
\footnote{This material was
presented in part at the rump session of the 14th Quantum Information Processing
(QIP) workshop, Singapore, Jan 2011.}}

\author{Naresh Sharma}
\email{nsharma@tifr.res.in}
\affiliation{School of Technology and Computer Science,
Tata Institute of Fundamental Research (TIFR), Mumbai 400 005, India}

\date{\today}

\begin{abstract}
It is known that repeated gambling over the outcomes of independent
and identically distributed (i.i.d.) random variables gives rise to  alternate
operational meaning of entropies
in the classical case in terms of the doubling
rates. We give a quantum extension of this approach for
gambling over the measurement outcomes of tensor product states.
Under certain parameters of the gambling setup, one can give operational
meaning of von Neumann entropies. We discuss two variants of
gambling when a helper is available and it is shown that the difference in their doubling rates is the quantum discord.
Lastly, a quantum extension of Kelly's gambling setup in the classical case
gives a doubling rate that is upper bounded by the Holevo information.
\end{abstract}

\maketitle

\section{Introduction}

Quantum information theory \cite{wilde-book,petz-book,nielsen-chuang} deals with the
information content in the quantum systems and is a generalisation of classical
information theory (see Ref. \cite{covertom} for example) for quantum systems.

The measurement outcome of a quantum system is a
random variable and the measurement alters the quantum state in general.
We confine ourselves to finite dimensional Hilbert spaces that
describe the quantum states and a probability mass function would describe the
measurement outcome
random variable that can be computed using the postulates of quantum mechanics.

If after a measurement, a quantum system is prepared again in the same
state as before the measurement and the same measurement process is repeated,
the sequence of the measurement outcomes is a sequence of i.i.d. random variables.

As an example, consider a quantum system prepared each time before the measurement
in the quantum state $\rho = p \ket{0} \bra{0} + (1-p) \ket{1} \bra{1}$, where
$0 \leq p \leq 1$. The measurement
operators are $\{\ket{0} \bra{0}, \ket{1} \bra{1}\}$. The measurement outcomes
form a sequence of i.i.d. binary random variables each of which take values $0$, $1$ with
probabilities $p$, $1-p$ respectively.

A classical gambling device such as a roulette consists of a revolving wheel onto
which a ball is dropped and the ball settles down to one of the numbered
slots or compartments on the wheel. Alice, the roulette player, bets on
a number or a subset of numbers on which the ball comes to rest. There is
a probability associated with winning on each gamble.

If bets are placed on the measurement outcomes of a quantum system,
then the apparatus becomes a gambling device or a quantum
roulette. Quantum gambling has been studied before in different contexts. Goldenberg
\emph{et al} invented a zero-sum game where a player can place bets at a casino located in
a remote site \cite{goldenberg-1999}. Hwang \emph{et al} considered its extensions using
non-orthogonal and more than $2$ states \cite{hwang-2001,hwang-2002}. Betting
on the outcomes of measurements of a quantum state was considered by
Pitowsky \cite{pitowsky-2003}.

We note that none of the above references study the log-optimal gambling strategies, which,
on the other hand, have been well-studied for the classical case (see for example
\cite{kelly-1956, covertom,erkip-cover-1998} and references therein).

Quantum systems exhibit certain characteristics that are not possible classically.
Bell inequalities \cite{bell-ineq-1964, chsh-ineq} give classical
limits to the figure of performance for certain setups and these
inequalities could be violated by quantum systems. We show that the quantum
gambling devices too exhibit certain characteristics that are impossible to replicate
classically.

At an information-theoretic level, the von Neumann entropy of the composite
quantum systems $A$ and $B$ can be smaller than the von Neumann entropy of
the subsystem $B$ alone giving rise to negative conditional entropies. This is,
as is well known, impossible for classical Shannon entropies.

Kelly defined a log-optimal gambling strategy by applying the law of large numbers
to the factor (a random variable)
by which Alice's wealth grows in a gamble. Thus, one can loosely claim that Alice's
wealth is an exponential function of the number of gambles \cite{kelly-1956}.
(We define this more precisely later.) This approach has been developed further with
the side information (or a helper) in Ref. \cite{covertom}.

The exponent (or the doubling rate if the base of the logarithm is $2$)
is a function of payoffs that the casino owner, Charlie, offers for each outcome,
outcome probability distribution, and Alice's strategy. When we optimise the strategy
under certain conditions, then the entropy (Shannon or von Neumann) appears in
the exponent.

We note that these entropies (and certain information measures) have deep
operational interpretations in classical and quantum information theory (see Ref.
\cite{wilde-book} and references therein).

In the classical case, Alice chooses how the wealth with which she is gambling
is going to be distributed across the various outcomes. As an example for two
outcomes, Alice could bet half of her money on each of the outcomes.

For the quantum case, Alice
can additionally make a choice of the measurement operators. Any classical
roulette would be a special case of a quantum roulette.

We also consider a case when a helper
named Bob is available for the gambler to make more money. Bob has access to
a quantum system that is correlated with Alice's quantum system. Bob is broke and
has no money to gamble on his system and offers Alice help in two ways.

In the first variant, Bob reports the measurement outcome to Alice
who now knows the collapsed state of her quantum system and uses this information
to further optimise her exponent (or the doubling rate). Alice may or may not
have control over the measurement operators applied by Bob.

In the second variant, Bob leases out his quantum system to Alice who then
gambles on the composite quantum system consisting of her and Bob's systems.
In return, Bob demands a share in Alice's accrued wealth and wants Alice to retain
the portion of wealth that Alice would have accrued by gambling only on her
system and had completely ignored the correlations
between the two systems. Bob's argument is that Alice can win more by taking the
correlations into account and it thus a win-win situation for both him, since he earns
money after being broke, and Alice since she earns more money.

Under certain conditions, for the classical gambling,
these two variants give rise to the same doubling rates whereas, for quantum gambling, they give rise to different
doubling rates whose difference is equal to the quantum discord,
a quantity that has been studied in a completely
different context \cite{verdal-2001,zurek-2002}.

Quantum discord is interpreted as purely the quantum part of the total
correlations between the two quantum systems. That these two variants
are the same classically in terms
of the doubling rate lends support to the above interpretation.

Kelly gave another interesting interpretation of the mutual information
\cite{kelly-1956}. Suppose Alice, knowing the output of a communication channel,
bets on the inputs to the channel, then under certain conditions, the optimised
doubling rate is equal to the mutual information. If one extends Kelly's result for
the quantum case, one gets a doubling rate in terms of a certain mutual
information that is a function of the measurement
operators and, using the Holevo bound, is upper bounded by the Holevo information.

We define a few quantities that will be needed later. The discrete or
the Shannon entropy of a random variable $A$ with probability
mass function $\pmb{p}^A$ is given by
$H(A)_{\pmb{p}}$ $\equiv -\sum_{i=1}^n p_i^A \log\left(p_i^A\right)$.
The classical relative entropy from $\pmb{p}$ to $\pmb{q}$
is given by $D\left(\pmb{p} || \pmb{q}\right)$ $\equiv \sum_i p_i \log ( p_i /q_i )$.
The von Neumann entropy of system $A$ in state $\rho^A$ is given by
$S(A)_\rho \equiv - \tr \rho^A \log \rho^A$. For a composite system $A$ and $B$
in state $\rho^{AB}$, the quantum conditional entropy of $A$ given $B$ is
$S(A|B)_{\rho} \equiv S(A,B)_\rho - S(B)_\rho$, and the quantum
mutual information is given by $S(A:B)_\rho \equiv S(A)_\rho - S(A|B)_\rho$.

\section{Rules of gambling}

Let us assume that there are $n$ outcomes of a gambling device.
Charlie decides that the payoff for the $i$th outcome is $o_i$-for-$1$, $i=1,...,n$.
In other words, if Alice puts down one dollar on outcome $i$ before the gamble,
she gets $o_i$ dollars if the outcome is $i$, and gets $0$ dollars if the outcome
is not $i$. Alice is allowed to bet on several outcomes in a gamble.
There is one and only one winning outcome.
Alice receives the payoff from the winning outcome and the bets on other outcomes are lost.

Alice can optimise on how she distributes her wealth across the outcomes. She may
decide not to bet on some outcomes and to gamble only with a fraction of her money
after each gamble and retaining a fraction of money. For quantum gambling, Alice
could additionally have control over which measurement operators to use and hence,
has some control over the outcome probabilities.

We shall assume that the casino never shuts down and that Alice can gamble there
as many times as she wants to.
For the ease of presentation and analysis, we shall not impose the restriction (common
in casinos) that Alice must always gamble with more than a minimum amount of
money or that the money can be gambled with and won in integer multiples of the smallest
unit of prevailing currency.

\subsection{Gambling with a helper}

Bob is the helper and has access to a random variable $B$ that is correlated with
the outcome of Alice's roulette modelled as a random variable $A$. Bob is broke
and doesn't have money to gamble on $B$.

Let us take the example of the American roulette with $38$ slots in
one of which the ball must fall in. There are two zero slots denoted by $0$ and $00$,
and $36$ numbered slots from $1$ to $36$. The bets are placed on the roulette table
layout. One game could be defined as betting on single numbers between $1$ to $36$
(also called ``Straight-up inside" bet). Charlie pockets the money if the outcome is
any of two zeros. Another game could be betting on which of three dozens the outcome
falls under (also called ``Dozen outside" bet). The same rule as in the previous game
applies for zero outcomes.

If $B$ is unknown,
then let us assume that $A$ takes values in the set $\{00,0,1,...,36\}$ and all its
elements are equally likely
to occur. Let us assume that $B$ takes values $0$ or $1$ and if $B$ takes value $0$,
then $A$ takes values among the first $19$ elements each with probability $2/57$
and the last $19$ elements each with probability $1/57$, and
otherwise, $A$ takes values among the first $19$ elements each with probability
$1/57$ and the last $19$ elements each with probability $2/57$.
Alice could use this extra information (outcome of $B$)
from Bob to perhaps gamble better.

The dependence of two random variables in the classical world is carried over to
the quantum world by considering two quantum systems that have non-zero total
correlation quantified by quantum mutual information \cite{nielsen-chuang}.
The total correlation can be further broken into a classical correlation and a purely
quantum quantity called the quantum discord \cite{verdal-2001,zurek-2002}.

Let us consider a quantum system $A$ on which Alice gambles. Alice could have a
control over how she wants to distribute her wealth across the outcomes and/or
what measurement operators she uses. Let us consider a quantum system $B$ 
that has a non-zero total correlation with $A$.

In one variant, Bob provides the outcomes of the measurement on $B$
to Alice who uses it to gamble better. Alice may or may not have control over
the measurement operators applied to $B$.
We shall see that as long as the classical correlation
(see Ref. \cite{verdal-2001}) between $A$ and $B$ is nonzero,
Alice can gamble better (the sense in which better is defined will be discussed in
greater detail).

In another variant, Bob leases out his system $B$ to Alice who could gamble in a larger Hilbert space
consisting of both the Hilbert spaces of $A$ and $B$. In return, Bob could demand
a share in Alice's earnings. Under certain conditions and a reasonable demand by Bob
of a share in Alice's earnings, we shall see that
as long as the total correlation between $A$ and $B$ is nonzero,
Alice can gamble better.

\section{Optimisation criterion}

\label{sec::crit}

The optimisation criterion is described in this section. For the simplicity of presentation,
we shall define the criterion for the classical case and the criterion will be
extended to the quantum case later.
Let us consider a random variable $A$ that describes the outcome on which the
bets are placed. At the start of each gamble, $A$ has a distribution
\beq
\pmb{p}^A = [p_1^A,...,p_n^A],
\enq
where $n$ is the number of values that $A$ takes and $\Pr\{A = i\} = p_i^A$.

Let the payoff for the $i$th outcome be $o_i^A$-for-$1$. Let us
assume that Alice distributes her wealth according to the probability vector
\beq
\pmb{q}^A = [q_1^A,...,q_n^A], ~~ \sum_{i=1}^n q_i^A = 1 - q_0^A,
~ q_i^A \geq 0, ~ \forall ~ i=0,...,n,
\enq
i.e., she puts $q_i^A$ fraction of her wealth on the $i$th outcome, $i=1,...,n$,
and $q_0^A$ is the fraction of wealth that Alice retains and does not gamble with.
At the start of each gamble, $A$ is prepared in the same state and has the probability mass
function as $\pmb{p}^A$. At the end of a
gamble, the factor by which Alice's wealth increases is a random variable denoted by
$X^A$, that takes values $q_0^A + q_i^A o_i^A$ with probability $p_i^A$, $i=1,...,n$.

Let us assume that for the $j$th gamble, the outcome random variable is
denoted by $X^A_j$. After $K$ gambles, Alice's wealth will grow by a factor of
\beq
S_K^A = \prod_{j=1}^K X^A_j.
\enq
It follows from our preparation that $X^A_1, X^A_2, ..., X^A_K$
are i.i.d. Kelly applied the weak law of large numbers (see Ref. \cite{feller-book} for example)
to the logarithm of $X_j^A$ \cite{kelly-1956}, and it follows that that for any $\epsilon > 0$,
\beq
\lim_{K \rightarrow \infty}
\Pr\left[ \left| \frac{1}{K} \sum_{k=1}^K \log(X^A_k) - W_A \right| > \epsilon \right] = 0,
\enq
where
\beq
\label{eq_wa}
W_A = \langle X^A \rangle = \sum_{i=1}^n p_i^A \log\left(q_0 + q_i^A o_i^A \right)
\enq
is the doubling rate of Alice's wealth and we assume that the base of $\log$ is
$2$ throughout this paper. Hence, for large $K$,
\beq
S_K^A \stackrel{.}{=} 2^{KW_A},
\enq
where $a_K \stackrel{.}{=} b_K$ denotes that
\beq
\lim_{K \rightarrow \infty} \Pr \left[ \frac{1}{K} \log \left( \frac{a_K}{b_K} \right) \right]
= 0.
\enq
We shall assume that Alice wants to optimise the doubling rate $W_A$.
Such a strategy is also called the log-optimal strategy.

\section{Optimisation of the doubling rate}

\label{sec::opt}

The optimisation in the classical case is done over $\pmb{q}^A$. The quantum case
is discussed subsequently where Alice could additionally have control over the measurement
operators.

Kelly identified three regimes based on the payoffs (see Refs. \cite{kelly-1956,covertom})
that are
\beq
\sum_{i=1}^n \frac{1}{o_i} ~~~~
\begin{array}{ll}
< 1, & \mbox{Super-fair odds} \\
= 1, & \mbox{Fair odds} \\
> 1, & \mbox{Sub-fair odds}
\end{array}
\enq
Kelly argued that for fair and super-fair odds, Alice does not need to retain any wealth
and $q_0^A=0$. This follows by choosing (a suboptimal choice in general)
$q_i^A = 1/o_i^A$, $i=1,...,n$, $q_0^A = 1 - \sum_{i=1}^n 1/o_i^A$, and
ending after the gamble with $1 + (1 - \sum_{i=1}^n 1/o_i^A)$ times the wealth
before the gamble.
Since the wealth increases for super-fair odds and remains the same with fair odds
with this suboptimal strategy, wealth cannot decrease by smartly gambling with all the
money rather than retaining a part of it.

For fair and super-fair odds, we choose $q_0^A=0$ and rewrite $W_A$ in (\ref{eq_wa}) as
\beqa
\label{doubling-rate}
W_A & = & \sum_{i=1}^n p_i^A \log(q_i^A o_i^A).
\enqa
The optimum doubling rate is
\beqa
W_A^* & = & \max_{\pmb{q}^A} W_A \\
& = & \sum_{i=1}^n p_i^A \log\left(o_i^A\right) - H(A)_{\pmb{p}} -
\min_{\pmb{q}^A} D\left(\pmb{p}^A || \pmb{q}^A\right).
\enqa
Since $\min_{\pmb{q}^A} D\left(\pmb{p}^A || \pmb{q}^A\right) = 0$ is achieved
at $\pmb{q}^A = \pmb{p}^A$, hence, for $\pmb{q}^A = \pmb{p}^A$,
\beq
W_A^* = \sum_{i=1}^n p_i^A \log\left(o_i^A\right) - H(A)_{\pmb{p}}.
\enq
Clearly, Alice needs to have an estimate of the probabilities of the measurement
outcomes to optimise her wealth's doubling rate. The optimising wealth distribution
being equal to the probability distribution of the outcome is also called proportional
gambling \cite{covertom}.

It is easy to see that for uniform fair odds, i.e., $o_i^A = o$, $i=1,...,n$,
\beq
W_A^* + H(A)_{\pmb{p}} = \log(o).
\enq
This is known as the conservation theorem stated as the sum of the doubling rate
and entropy is constant for uniform fair odds \cite{covertom}. In this case, the low
entropy gambling devices result in larger doubling rates.

For sub-fair odds, Kelly \cite{kelly-1956} found the optimum solution as
\beq
W_A^* = \gamma D \left( \acute{\pmb{p}}^A || \pmb{\sigma}^A \right) +
D \left( [\gamma, 1-\gamma] ~ || ~ [\beta,1-\beta] \right),
\enq
where $\gamma = \sum_{i \in I} p_i$, $\acute{\pmb{p}}^A = \{p_i^A/\gamma\} \Big|_{i \in I}$,
$\beta  = \sum_{i \in I} 1/o_i^A$,
$\sigma_i^A = 1/ \left(\beta o_i^A \right)$,
$\pmb{\sigma} = \{\sigma_i^A\} \Big|_{i \in I}$, and $I$ is a subset of
$\{1,2,...,n\}$ that is uniquely determined
by $\beta < 1$, $p_i^A o_i^A > (1-\gamma)/(1-\beta)$ for $i \in I$,
$p_i^A o_i^A \leq (1-\gamma)/(1-\beta)$ for $i \notin I$. Optimisation can be
done using the Karush-Kuhn-Tucker conditions (see Ref. \cite{bertsekas} for example).

Clearly, the expression for the optimum solution is not simple for the case of sub-fair odds. 
In the rest of the paper, we shall compute the doubling rates for fair or super-fair odds and
not for the case of sub-fair odds. We note, however, that the protocols that we
give for quantum gambling are independent of the odds and would apply to sub-fair
odds as well.

\subsection{Optimisation in the presence of a helper}

\label{sec::opt-side}

Charlie, the casino owner, allows Alice to use a helper named Bob. Bob has access to a gambling device whose outcome random variable $B$ is
correlated to that of $A$ (defined in Section \ref{sec::crit}) and let their
joint probability mass function be $\Pr\{A=i,B=j\}$, $i=1,...,n$, $j=1,...,m$,
where $B$ takes $m$ values.

Bob reports the outcome of his gambling device, say $j$, to Alice
and the optimum strategy for Alice as per the discussion above for fair or super-fair odds
is to choose
\beq
q^{A|B}_{i|j} = \Pr\{A=i | B = j\} = \frac{ \Pr\{A = i, B = j\} } {p_j^B}, ~ i = 1,...,n,
\enq
where $p_j^B = \Pr\{B=j\}$.
With this choice, we get the doubling rate as
\beq
\label{cond-dr}
W_{A|B}^* = \sum_{i=1}^n p_i^A \log\left(o_i^A\right) -
\sum_{j=1}^m p_j^B  H(A)_{\pmb{p}^{A|B=j}},
\enq
where
\beq
\pmb{p}^{A|B=j} = \left[\Pr\{A=1 | B = j\}, ... , \Pr\{A=n | B = j\} \right].
\enq
We note that $\sum_{j=1}^m p_j^B  H(A)_{\pmb{p}^{A|B=j}}$ is the
conditional entropy of $A$ given $B$.

In another variant,
Bob leases out his gambling device to Alice allowing Alice to gamble on both
$A$ and $B$. In return, Bob demands that Alice retains $2^{-K W_B^*}$ fraction of
her earnings and the rest would be pocketed by him.
Bob's argument is that if Alice gambles right on $B$ by completely ignoring the
correlations, then Alice's wealth increases additionally by a factor of $2^{K W_B^*}$, and certainly
Alice can make more by exploiting the correlations.

It turns out, as the computation below shows, that for certain choices of the odds,
both the variants of the help that Bob offers turn out to be the same in terms of
the doubling rates. We rewrite $W_{A|B}^*$ from \eqref{cond-dr} as
\beqa
W_{A|B}^* & = & \sum_{i=1}^n \sum_{j=1}^m p_i^A p_j^B \log\left(o_i^A o_j^B\right) -
H(A,B)_{\pmb{p}} - \left[ \sum_{j=1}^m p_j^B \log\left(o_j^B\right) -
H(B)_{\pmb{p}} \right], \\
& = & W_{A,B}^* - W_B^*,
\enqa
where $\pmb{p}^{A,B}=[\Pr\{A=1,B=1\},...,\Pr\{A=n,B=m\}]$,
$W_{A,B}^*$ and $W_B^*$ are the optimum doubling rates of the fictitious games that
are played over systems with outcome probability distributions given by
joint distribution of $A$ and $B$, and $B$ alone respectively, and
$o_i^A o_j^B$-for-$1$ is the payoff when the
$(A,B)$ takes the value $(i,j)$ and $o_j^B$-for-$1$ is the
payoff when $B$ takes the value $j$.

As we shall see, these two variants don't give the same doubling rates for the
quantum gambling.

\section{Quantum gambling}

\label{sec:model}

Consider a quantum system $A$ that is described by a Hilbert space ${\mathcal{H}}_A$
of dimension $\dim(A)$. At the start of each gamble, Charlie prepares the
quantum system $A$ in state $\rho^A$.
Alice may not know this state but learns it over repeated gambling. We ignore the
time it takes for Alice to learn the state since we consider the number of times
Alice gambles, $K$, is large enough.

Clearly, there needs to be some
checks on Charlie and the National Gaming Commission sends its representatives
to Charlie's casino to ensure that Charlie is honest in stating that the state at the start
of each gamble is indeed $\rho^A$. These
representatives could employ a scheme described by Blume-Kohut and Hayden
in Ref. \cite{kohut-2006} for accurate quantum state estimation via
``keeping the experimentalist honest".

The measurement is completely
described by $n$ measurement operators $\{E_i^A; i=1,...,n\}$.
It seems reasonable to assume at this point that $n = \dim(A)$ and
the measurement operators are orthogonal projectors
($\left(E_i^A\right)^\dagger = E_i^A$,
$E_i^A E_j^A = \delta_{i,j} E_i^A$, where $\delta_{ij} = 1$ if $i=j$, and is $0$ otherwise)
to rule out the possibility that Alice could win each time unless $\rho^A$ has rank $1$.
The measurement operators are complete and hence,
$\sum_{i=1}^n E_i^A = \mathbbm{1}$, where $\mathbbm{1}$ is the
Identity matrix whose dimensions should be clear from the context.
The superscript $A$ is added to indicate the quantum system on which gambling is
carried out.

The probability of the $i$th outcome in the quantum roulette is given by
\beq
p_i^A = \tr \, \rho^A E_i^A
\enq
and we define $\pmb{p}^A = [p_1^A,...,p_n^A]$.

For the $i$th outcome, the quantum state collapses to $E_i^A \rho^A E_i^A /p_i^A$
after the measurement. We shall assume that for the next gamble, the state is again
prepared to be $\rho^A$ and the measurement using $\{E_i^A\}$ is applied again.
It now follows easily from the treatment in Section \ref{sec::opt} that
\beq
W_A^* = \sum_{i=1}^n p_i^A \log\left(o_i^A\right) - H(A)_{\pmb{p}}.
\enq
Next, we consider the case of uniform fair odds, $o_i^A = o$ for all $i$,
where Alice has control over both $\pmb{q}^A$ and
$\{E_i^A\}$. The additional choice of measurement operators distinguishes
classical and quantum gambling. The optimum doubling rate is given by
\beq
W_A^{**} = \max_{\{E_i^A\}} \max_{\pmb{q}^A} W_A.
\enq
It follows from the convexity of $t \mapsto t\log(t)$ that
\beqa
H(A)_{\pmb{p}} & \geq & S(A)_\rho,
\enqa
and the equality is achieved if and only if $E_i^A$ are the eigenvectors
of $\rho^A$. Hence, it follows that
\beqa
\label{flag-1}
W_A^{**} & = & \log(o) - S(A)_\rho.
\enqa
We note here that the doubling rate is larger in general if Alice optimises over both
$\pmb{q}^A$ and $\{E_i^A\}$ rather than just optimising over $\pmb{q}^A$ alone.

We shall add superscript $*$ on the doubling rates if Alice optimises over the wealth
distribution and superscript $**$ if Alice optimises over the wealth distribution as well as the
measurement operators.

\section{Quantum gambling with a helper}

Bob is the helper with access to a quantum system $B$ described by Hilbert space
${\mathcal{H}}_B$ and at the start of each gamble,
the joint state in the composite system $AB$ described by
${\mathcal{H}}_A \otimes {\mathcal{H}}_B$ is $\rho^{AB}$. We shall
follow the convention that
$\rho^A = \tr_B \, \rho^{AB}$ and $\rho^B = \tr_A \, \rho^{AB}$.
As mentioned before, we assume that Bob is broke
and doesn't have money to gamble on $B$.
We look at two ways in which Bob renders help.

\subsection{Bob reports the outcomes to Alice}

The protocol is described as follows:

\begin{enumerate}
\item Bob measures the system $B$ using a complete set of measurement operators
$\{F_j^B\}$, $j=1,...,m$, $\sum_{j=1}^m \left(F_j^B\right)^\dagger
F_j^B = {\mathrm{I}}$. Note that apriori we don't place any restrictions on
$\{F_j^B\}$ and the measurement may not be a projective.
Alice may or may not have control over $\{F_j^B\}$.
For the $j$th outcome in $B$, the state of system $A$ collapses to
\beq
\rho^A_j = \frac{1}{\beta_j}
\tr_B \, \rho^{AB} \left[ {\mathrm{I}} \otimes \left( F_j^B \right)^\dagger F_j \right],
\enq
where
\beq
\beta_j = \tr \, \rho^{AB} \left[ {\mathrm{I}} \otimes \left(F_j^B\right)^\dagger
F_j \right]
\enq
is the probability of the $j$th outcome of measurement in system $B$. Bob tells the
measurement outcome to Alice.

\item Alice uses the measurement outcome in $B$ to distribute her wealth
across the
outcomes. If Alice has control over the measurement operators as well, Alice could
tune these measurement operators depending on the measurement outcome in $B$
and uses $\{E_{ij}^{A}\}$
(projective measurement) for the $j$th outcome in system $B$. If Alice has no control
over the measurement operators, then $E_{ij}^A = E_i^A$, $j=1,...,m$.

\end{enumerate}

The probability of the $i$th outcome in system $A$ given the $j$th
outcome in system $B$ is given by
\beqa
\alpha_{i|j}^A & = & \tr \, \rho^A_j E_{ij}^A \\
& = & \frac{1}{\beta_j}
\tr \, \rho^{AB} \left( E_{ij}^A \otimes \left(F_j^B\right)^\dagger F_j^B \right)
\enqa
and let $\pmb{\alpha}_{(j)}^A = \left[ \alpha_{1|j}^A,...,\alpha_{n|j}^A\right]$.

We first consider the case where Alice does not control the measurement operators.
For this case, $E_{ij}^A = E_i^A$, i.e., a fixed set of measurement operators is applied for
each outcome $j$ in system $B$, and
\beq
\pmb{p}^A = \sum_{j=1}^m \beta_j \pmb{\alpha}_{(j)}^A,
\enq
where $\pmb{p}^A$ is the probability vector that gives the probability of outcomes
in system $A$ without Bob's help.
The optimal doubling rate given the $j$th outcome in system $B$ is
\beq
W_{A|B,j}^* = \sum_{i=1}^n \alpha_{i|j}^A \log\left(o_i^A\right) -
H(A)_{\pmb{\alpha}_{(j)}^A}.
\enq
The overall doubling rate is given by
\beqa
W_{A|B}^* & = & \sum_{j=1}^m \beta_j W_{A|B,j}^* \\
& = & \sum_{i=1}^n \sum_{j=1}^m \beta_j \alpha_{i|j}^A \log\left(o_i^A\right)
- \sum_{j=1}^m \beta_j H(A)_{\pmb{\alpha}_{(j)}^A}.
\enqa
For uniform fair odds, the increase in doubling rate due to Bob's help is
\beq
W_{A|B}^* - W_A^* = H\left(\pmb{p}^A\right) - \sum_{j=1}^m
\beta_j H(A)_{\pmb{\alpha}_{(j)}^A},
\enq
which, from the concavity of the entropy, is always nonnegative.

Next, we consider the case where Alice has control over the measurement
operators. We note that in this case, $W_{A|B}-W_A$ (note that these are not
optimised) can be negative (since
$\pmb{p}^A \neq \sum_{j=1}^m \beta_j \pmb{\alpha}_{(j)}^A)$, which is
impossible in a classical gambling. As an example, consider uniform fair odds and
let $\dim(A) = \dim(B) = 2$,
$\rho^{AB} = \ket{0}\bra{0} \otimes \rho^B$. While quantum systems $A$ and $B$ in
this state have no correlation, but it still allows us to give a simple example
to prove the above point.
Let the measurement operators for $A$ without the help
be $\{\ket{0}\bra{0}, \ket{1}\bra{1}\}$. Let the measurement outcome of $B$
be $0$ or $1$ with probability $0.5$ each. If outcome in $B$ is $0$,
then the measurement operators for $A$ are $\{\ket{0}\bra{0}, \ket{1}\bra{1}\}$
and if the outcome in $B$ is $1$, then the measurement operators for $A$ are
$\{ \ket{-}\bra{-}, \ket{+}\bra{+}\}$, where $\ket{-} = \{(\ket{0}-\ket{1})/\sqrt{2}$ and
$ \ket{+} = (\ket{0}+\ket{1})/\sqrt{2}$.
This choice gives $H(A)_{\pmb{p}} = $ $H(A)_{\pmb{\alpha}_{(1)}^A} = 0$,
and $H(A)_{\pmb{\alpha}_{(2)}^A} > 0$.

The increase in the optimum doubling rate is given by
\beqa
W_{A|B}^{**} - W_A^{**} & = & \max_{\{E_{ij}^A\}}
\left[ \sum_{i=1}^n \sum_{j=1}^m \beta_j \alpha_{i|j}^A \log\left(o_i^A\right)
-  \sum_{j=1}^m \beta_j H(A)_{\pmb{\alpha}_{(j)}^A} \right] \nonumber \\
& & ~~~~~ - \max_{\{E_{i}^A\}}
\left[ \sum_{i=1}^n p_i^A \log\left(o_i^A\right) - H(A)_{\pmb{p}} \right].
\enqa
If Alice has control over $\{F_j^B\}$ as well,
it follows for uniform fair odds and from (\ref{flag-1}) that
\beq
W_{A|B}^{**} - W_A^{**} = \max_{\{F_j^B\}} \left[ S(A)_\rho
- \sum_{j=1}^m \beta_j S(A)_{\rho_j} \right].
\enq
We note that this quantity is referred to as
the classical correlation between the quantum systems $A$ and $B$ \cite{verdal-2001}.

Suppose Alice is given a choice where Bob either measures the system $B$ or
applies a quantum operation ${\cal F}(\cdot)$ on $B$, and then Alice gambles on system $A$.
For uniform fair odds with system $A$ measured in its eigenbasis,
measuring $B$ is a better option. To see this, note that
the state of $A$ after the operation on $B$ is given by
\beq
{\cal F}(\rho^A) = \tr_B \, \sum_{j=1}^m \left[{\mathrm{I}} \otimes F_j^B \right]
\rho^{AB} \left[{\mathrm{I}} \otimes \left( F_j^B\right)^\dagger \right],
\enq
where $\{F_j^B\}$ are the Kraus operators characterising ${\cal F}(\cdot)$. Invoking
the concavity of the von Neumann entropy and assuming that $A$ is measured in
its eigenbasis, we get
\beq
\label{cl_corr}
S(A)_{ {\cal F}(\rho^A) } \geq \sum_{j=1}^m \beta_j S(A)_{\rho_j},
\enq
where $\rho^A_j$ and $\beta_j$ are the same as above. Hence, in this case,
it is better to measure the state $B$ rather than apply a quantum operation on $B$.

\subsection{Bob leases out $B$ to Alice}

\label{protocol22}

The protocol is described as follows:

\begin{enumerate}

\item Bob leases $B$ to Alice that gives her the freedom that she can gamble on the
composite system $AB$ instead of $A$ alone.

\item In return of this offer, Bob demands that at the end of
$K$ gambles, Alice can keep $2^{-KW_B}$ fraction of her money and the rest will be kept
by Bob, where $W_B$ is an achievable doubling rate Alice would get if she had gambled on
$B$ alone. Note that Bob could choose $W_B = W_B^*$ or $W_B = W_B^{**}$. The choice
of this fraction in the lease agreement has the same motivation as in the classical case 
in Sec. \ref{sec::opt-side}.

\end{enumerate}

Let us assume that the measurement operators for the composite system $AB$
are given by $G_{ij}^{AB}$, $i=1,...,n$, $j=1,...,m$. Let the payoff for the
output $(i,j)$ be $\left(o_i^A o_j^B\right)$-for-$1$. (Charlie could choose dependent odds for $AB$ as well.) Let
$p_{ij}^{AB} = \tr \, \rho^{AB} G_{ij}^{AB}$ and
$\pmb{p}^{AB} = [p_{11}^{AB},...,p_{1,m}^{AB},...,p_{n1}^{AB},...,p_{nm}^{AB}]$.
Let the measurement operators for $B$ be given by $\{F_j^B\}$, $j=1,...,m$,
$p_j^B = \tr \, \rho^{AB} \left({\mathrm{I}} \otimes F_j^B \right)$, and
$\pmb{p}^B = [p_1^B,...,p_m^B]$. Let Alice bet $q_{ij}^{AB}$ fraction of her
wealth on the outcome $(i,j)$.

After accounting for Bob's share (computed using $W_B = W_B^*$),
the doubling rate for Alice's wealth is given by
\beqa
W_{A|B} & = & \sum_{i=1}^n \sum_{j=1}^m p_{ij}^{AB} \log\left(q_{ij}^{AB}
o_i^A o_j^B\right) - W_B^*,
\enqa
where
\beq
W_B = \sum_{j=1}^m p_j^B \log(o_j^B) - H(B)_{\pmb{p}^B}.
\enq
We note that in this case, in general, $p_j^B \neq \sum_{i=1}^n p_{ij}^{AB}$. First
consider the case where Alice can only control the wealth distribution and not the
measurement operators.
As per our discussion before, the optimal thing for Alice to do would be to choose
$q_{ij}^{AB} = p_{ij}^{AB}$ to get
\beq
\label{dummy1}
W_{A|B}^* = \sum_{i=1}^n \sum_{j=1}^m p_{ij}^{AB} \log\left( o_i^A o_j^B\right)
- \sum_{j=1}^m p_j^B \log\left(o_j^B\right)  - H(A,B)_{\pmb{p}^{AB}} +
H(B)_{\pmb{p}^B}.
\enq

If Alice also controls the measurement operators, then if Bob chooses
$W_B=W_B^{**}$ to compute his share as per the protocol, the doubling rate
for Alice's wealth is given by
\beq
W_{A|B}^{**} = \max_{\{ G_{ij}^{AB} \}} \left[ \sum_{i=1}^n \sum_{j=1}^m
p_{ij}^{AB}
\log\left( o_i^A o_j^B\right) - H(A,B)_{\pmb{p}^{AB}} \right]
- \max_{\{F_j^B\}} \left[
\sum_{j=1}^m p_j^B \log\left(o_j^B\right) - H(B)_{\pmb{p}^B} \right].
\enq
In the special case of uniform fair
odds of $o$-for-1 for both $A$ and $B$, we get
\beqa
W_{A|B}^{**} & = & \log(o) - S(A,B)_\rho + S(B)_\rho \\
\label{neg_cond}
& = & \log(o) - S(A|B)_\rho.
\enqa
Unlike the classical case, the quantum conditional entropy can be negative and can
result in attaining the doubling rates in quantum gambling that cannot be achieved
by any classical gambling.

As an example, consider a Bell pair
\beq
\ket{\beta_{00}} = \frac{\ket{00} + \ket{11}}{\sqrt{2}},
\enq
and $\rho^{AB} = \ket{\beta_{00}} \bra{\beta_{00}}$. Both $A$ and $B$
are described by a Hilbert space of dimension $2$. In this case,
$S(A,B)_\rho = 0$
since $AB$ is in a pure state, but $\rho^B = {\mathrm{I}}/2$, and hence,
$S(B)_\rho = 1$. So, $S(B|A)_\rho = -1$.

Alice computes the change in the doubling rate of her take-home income as
\beqa
W_{A|B} - W_A & = & \sum_{i=1}^n \sum_{j=1}^m p_{ij}^{AB}
\log\left( o_i^A o_j^B \right) - \sum_{j=1}^m p_j^B \log(o_j^B)  -
\sum_{i=1}^n p_i^A \log(o_i^A) \nonumber \\
& & ~~~~
- H(A,B)_{\pmb{p}^{AB}} + H(B)_{\pmb{p}^B} + H(A)_{\pmb{p}^A},
\enqa
where $p_i^A$ is the probability of the $i$th outcome in $A$ and
$\pmb{p}^A = [p_1^a,...,p_n^A]$.
For $W_B = W_B^{**}$, uniform fair odds of $o$-for-$1$,
and with Alice having control over the measurement operators, and
the systems $A$, $B$,
and $AB$ being measured in their respective eigenbases, and we get
\beq
\label{qu_mi}
W^{**}_{A|B} - W^{**}_A = S(A : B)_\rho,
\enq
where $S(A:B)_\rho \geq 0$ for all $\rho^{AB}$. So, if Alice
gambles right, she can still make money or not lose money (if $S(A:B)=0$)
despite Bob demanding a share in Alice's earnings.

\subsection{Discussion}

We first note that although both the variants of using the helper are
the same for classical gambling, they are, in general, different for quantum gambling.

The difference in the optimum doubling rates
by using the helper in the above two variants under uniform fair odds,
systems $A$ and $B$ being measured in their respective eigenbases, Bob demanding
that Alice can only keep $2^{-KW_B^{**}}$ of her earnings in $K$ gambles
is given by
\beqa
D(A \rangle B)_\rho & = & W^{**}_{A|B} \Bigg|_{\text{Variant~2}}
- W^{**}_{A|B}\Bigg|_{\text{Variant~1}} \\
& = & S(A:B)_\rho - \max_{\{F_j\}} \left[ S(A)_\rho
- \sum_{j=1}^m \beta_j S(A)_{\rho^A_j} \right],
\enqa
and is called the quantum discord between $A$ and $B$ (see \cite{verdal-2001,zurek-2002}
and the citing articles) and has been studied in a completely different context.
This quantity is nonnegative and is zero for the classical case.
It is shown by Ferraro \emph{et al} that with probability 
one, a quantum state chosen at random would have a non-zero quantum discord
\cite{ferraro-2010}.

\subsection{Alternating quantum system of the helper}

Charlie may choose to concoct a different protocol to take care of the negative conditional
entropy in (\ref{neg_cond}). Charlie prepares the state for each gamble to be
$\rho^{ABC}$ in the composite system $ABC$ described by the Hilbert space
${\mathcal{H}}_A \otimes {\mathcal{H}}_B \otimes {\mathcal{H}}_C$.
Alice gambles on $A$ as before while Bob has access to $B$
in $f$ fraction of the gambles and to $C$ in
$(1-f)$ fraction of the gambles and Bob knows whether the system he has access
to on that gamble is $B$ or $C$. Bob takes his share of Alice's earnings in both the
cases as described in Section \ref{protocol22}. It follows that for uniform fair odds of
$o$-for-$1$ for $A$, $B$, and $C$, systems $A$, $B$, and $C$ measured in their
respective eigenbases, Bob computing his share by choosing $W_B = W_B^{**}$ and
$W_C = W_C^{**}$, and assuming that $fK$ is an integer
($K$ is the number of gambles), the doubling rate is given by
\beq
W^{**}_{A|BC} = \log(o) - f S(A|B)_\rho - (1-f) S(A|C)_\rho.
\enq
It is clear that there would exist an $f$ that would do the job since
by choosing $f=1/2$ and invoking strong sub-additivity
\cite{lieb-ruskai-ssa0-1973,lieb-ruskai-ssa-1973}, we get 
\beq
f S(A|B)_\rho + (1-f) S(A|C)_\rho \geq 0.
\enq

\subsection{Quantum extension of Kelly's setup}

Kelly defined another scenario wherein the gambler on receiving
the output of a channel bets on the input that was transmitted \cite{kelly-1956}. Under
certain parameters, Kelly showed that the doubling rate is equal to
the mutual information between the input and the output of the channel, whose
maximum over the input probability distribution, as is well known,
is the capacity of the channel \cite{covertom}.

It is not difficult to present a quantum extension of Kelly's setup as follows.
We first discuss Kelly's classical setup.

Let $p_i$ be the probability of $i$th input to the channel, $p_{j|i} = \Pr\{ \text{Output} = j |
\text{Input} = i\}$, and $o_i$-for-$1$ are the odds
for the $i$th input where $o_i = 1/p_i$. It is not difficult to
show that the maximum achievable doubling rate
for fair or super-fair odds Alice can achieve is by betting $q_{i|j} = \Pr\{\text{Input} = i|
\text{Output} = j\}$ fraction of her wealth on the $i$th input given that the output is $j$
and is given by the mutual information between the input and the output.

The doubling rate from Eq. \eqref{doubling-rate} for a given measurement outcome
$j$ is given by
\begin{align}
W_j & = \sum_i q_{i|j} \log\left(q_{i|j} o_i \right) \\
& = \sum_i q_{i|j} \log\left(\frac{q_{i|j}}{p_i} \right),
\end{align}
and the overall doubling rate is
\beq
W = \sum_j q_j W_j = \sum_{i,j} q_j q_{i|j} \log\left(\frac{q_{i|j}}{p_i} \right),
\enq
which is the mutual information between the input and the output and $q_j = \Pr\{\text{Output} =j\}$.

A quantum extension of Kelly's setup is as follows. For the $i$th input that is chosen with
probability $p_i$, a density matrix $\rho_i^A$ is generated and is measured using the POVM
$\{\Lambda_j^A\}$. Hence the probability of the $j$th outcome given $i$th input is
\beq
p_{j|i} = \tr \, \rho_i^A \Lambda_j^A.
\enq
Continuing with Kelly's optimisation, we get the doubling rate as the mutual
information, which using the Holevo's bound (see Refs. \cite{wilde-book, nielsen-chuang}
for example) is upper bounded by the Holevo information given by
\beq
\chi(\{p_i,\rho_i\}) = S(A)_{\sum_i p_i \rho_i^A} - \sum_i p_i S(A)_{\rho_i^A}.
\enq

\subsection{Further variants}

We note here that several other variants can be concocted. For example,
one could put a classical communication channel with helper's information as
the input and Alice receiving the output. Alice receives the noisy information
and could process it and then use it for gambling.

In another variant, Alice could
choose to first apply one of the available (as provided by Charlie) quantum
operations on the state and then gamble on the new state (see also
Ref. \cite{goldenberg-1999}).

It should be apparent that analysis of such protocols can be done on a specific
choice of the channel in the former case and the available restricted quantum
operations provided by Charlie in the latter case.

\section{Conclusions}

We studied the
log-optimal strategies for some quantum gambling protocols.
We first considered the case where Alice gambles on a quantum system (or roulette)
by distributing her wealth across the outcomes and/or by choosing the measurement
operators. Next, we considered the case where Charlie allows Alice to have a helper
Bob with access to another quantum system that is correlated with Alice's quantum roulette and
considered two variants. In one variant, Bob reports the measurement outcome on his
system to Alice who uses it to gamble better. In another variant, Alice gambles on the composite
system consisting of her quantum roulette and Bob's system and Bob takes a
pre-specified cut in Alice's wealth in return.
The difference in the doubling rates of these two variants is the quantum discord, which
is zero for classical roulettes. 
We also considered quantum extension of Kelly's setup.
Finally, we considered the case of alternating quantum system of the helper.
Quantum gambling can have purely quantum effects that cannot be replicated by any
classical roulette.

\bibliographystyle{unsrturl}
\bibliography{master}

\end{document}